\documentclass[%
 aip,
 amsmath,amssymb,
preprint,%
]{revtex4-1}

\usepackage{CJKutf8}
\usepackage{graphicx}
\usepackage{dcolumn}
\usepackage{bm}

\usepackage[utf8]{inputenc}
\usepackage[T1]{fontenc}
\usepackage{mathptmx}
\usepackage{etoolbox}
\usepackage{float}
\usepackage{microtype}
\usepackage{graphicx}
\usepackage{subfigure}
\usepackage{booktabs} 

\usepackage{amsmath}
\usepackage{amssymb}
\usepackage{mathtools}
\usepackage{amsthm}
\usepackage{commath}

\usepackage{hyperref}
\usepackage{tabularx}
\usepackage{multirow}
\usepackage[table,dvipsnames]{xcolor}
\usepackage{tikz,siunitx}

\usepackage[capitalize]{cleveref}

\makeatletter
\def\@email#1#2{%
 \endgroup
 \patchcmd{\titleblock@produce}
  {\frontmatter@RRAPformat}
  {\frontmatter@RRAPformat{\produce@RRAP{*#1\href{mailto:#2}{#2}}}\frontmatter@RRAPformat}
  {}{}
}%
\makeatother
\begin{document}

\begin{CJK*}{UTF8}{gbsn}

\preprint{AIP/123-QED}
\title{Multi-fidelity Deep Learning-based methodology for epistemic uncertainty quantification of turbulence models}

\author{Minghan Chu}
\email{17mc93@queensu.ca.}
 \affiliation{Mechanical and Materials Engineering Department, Queen's University, Kingston, ON K7L 2V9, Canada.}
 
\author{Weicheng Qian}%
\affiliation{ 
Department of Computer Science, University of Saskatchewan, Saskatoon, SK S7N 5C9 Canada.
}%

\date{\today}

\date{\today}

\begin{abstract}

Computational Fluid Dynamics (CFD) simulations using turbulence models are commonly used in engineering design. Of the different turbulence modeling approaches that are available, eddy viscosity based models are the most common for their computational economy. Eddy viscosity based models utilize many simplifications for this economy such as the gradient diffusion and the isotropic eddy viscosity hypotheses. These simplifications limit the degree to which eddy viscosity models can replicate turbulence physics and lead to model form uncertainty. The Eigenspace Perturbation Method (EPM) has been developed for purely physics based estimates of this model form uncertainty in turbulence model predictions. Due to its physics based nature, the EPM weighs all physically possible outcomes equally leading to overly conservative uncertainty estimates in many cases. In this investigation we use data driven Machine Learning (ML) approaches to address this limitation. Using ML models, we can weigh the physically possible outcomes by their likelihood leading to better calibration of the uncertainty estimates. Specifically, we use ML models to predict the degree of perturbations in the EPM over the flow domain. This work focuses on a Convolutional Neural Network (CNN) based model to learn the discrepancy between Reynolds Averaged Navier Stokes (RANS) and Direct Numerical Simulation (DNS) predictions. This model acts as a marker function, modulating the degree of perturbations in the EPM. We show that this physics constrained machine learning framework performs better than the purely physics or purely ML alternatives, and leads to less conservative uncertainty bounds with improved calibration. 
\end{abstract}

\maketitle
\end{CJK*}

\section{\label{sec:level1}INTRODUCTION}


Turbulent flows are important in various applications involving engineering design. These problems can range from aerospace applications like the fluid flow over an aircraft's wing or through an engine diffuser, astrophysical applications like investigation of the Jovian atmosphere, biomedical applications like the design of cardiovascular devices, etc. Turbulence models for such diverse applications need to be accurate across these applications. Additionally these models need to be computationally inexpensive due to computational cost considerations where engineering design is an iterative process. Reynolds Averaged Navier Stokes (RANS) models are commonly used in these CFD applications. The most commonly used RANS turbulence models are eddy viscosity based turbulence models. They are the workhorse for engineering design applications. Eddy viscosity models utilize many simplifications for their computational economy such as the gradient diffusion and the turbulent viscosity hypotheses (TVH). These simplifications limit the degree to which eddy viscosity models can replicate turbulence physics especially in complex real life turbulent flows like those with turbulent separation and reattachment, surface curvature, etc\cite{craft1996development}. As using models with substantial predictive uncertainty can lead to sub-optimal designs with limited efficiency and reliability, reliable estimates of predictive uncertainty are required to redress this scenario.

Uncertainties in the CFD simulation predictions using turbulence models arise from two reasons, aleatoric and epistemic uncertainties\cite{smith2013uncertainty}. Aleatoric uncertainties are introduced from the imprecision of a system \cite{duraisamy2019turbulence}. Sources of aleatoric uncertainty include differences in the initial conditions or boundary conditions between simulations and the real life experiments, variations in geometry and material properties from errors in measurement, etc. Aleatoric uncertainties are referred to as irreducible or stochastic uncertainty. Therefore investigations focusing on aleatoric uncertainties re-formulate the problem in stochastic terms using a probabilistic formulations that utilizes random variables. This includes the use of Polynomial Chaos \cite{najm2009uncertainty} and Stochastic Collocation \cite{Mathelin} based approaches, expressing closure coefficients as random variables \cite{loeven2008airfoil,ahlfeld2017single}, expressing flow domain as a stochastic field \cite{dow2015implications,doostan2016bi}, characterizing the initial and boundary conditions as stochastic \cite{pecnik2011assessment}, etc. 

Separate from the stochastic nature of aleatoric uncertainty that may add unbiased noise to the predictions, epistemic uncertainties are caused by incomplete or imperfect representation of turbulence physics and add bias to model predictions. Epistemic uncertainty in turbulence models are caused due to our limitations in our understanding of turbulent flow physics and the inability to express the physics of turbulence in models. Epistemic uncertainty includes parameter uncertainty associated with the coefficients of the turbulence closure model and the model-form or structural uncertainty associated with the limitations of the model expression. Of these the model form uncertainties can be a dominant source of uncertainty in complex real life turbulent flows of engineering interest \cite{duraisamy2017status}. To reduce model form uncertainties in turbulence simulations investigators may choose to employ higher fidelity models, for example selecting the use of Large Eddy Simulations (LES) based simulations instead of an eddy viscosity model. But this refinement in model fidelity comes with a steep increase in computational expense of many orders of magnitude. This additional computational cost is impractical for engineering design and analysis. Therefore it is important to retain the computational economy of eddy viscosity models while also being able to quantify their model form uncertainties.  

The Eigenspace Perturbation Method (EPM) \cite{iaccarino2017eigenspace} is the only physics based approach to quantify turbulence model form uncertainties. This method uses tensor perturbations in the Spectral decomposition components of the predicted Reynolds stresses to estimate the sensitivity of the predictions. The range of the perturbed predictions is used to assess the prediction intervals on the Quantities of Interest. Due to its efficacy and low computational cost this method has been applied across different fields of engineering in the past including the design of urban canopies\cite{gorle2019epistemic}, aerospace design and analysis  \cite{mishra2019uncertainty, mishra2017rans, mishra2019estimating, mishra2017uncertainty}, application to design under uncertainty (DUU) \cite{demir2023robust, cook2019optimization, mishra2020design, righi2023uncertainties}, virtual certification of aircraft designs \cite{mukhopadhaya2020multi, nigam2021toolset}, etc . 

The central limitation of the EPM is that it assumes the worst case situation and thus leads to overly conservative uncertainty estimates. The magnitude of perturbations in the EPM are reflective of the degree of discrepancy in the model prediction. Turbulence model discrepancy is not uniform over the entire flow domain but is varying, with a higher discrepancy in region with sharp curvature of flow streamlines, flow separation, flow re-attachment, separation bubbles, etc. The perturbations should also reflect this spatial variation in the discrepancy by varying their magnitude. This can be carried out using a ``marker" function that estimates the discrepancy in the turbulence model predictions over the flow domain and modulates the degree of the perturbations accordingly. However there is no purely physics based marker function that can ascertain this spatial variation in perturbations. This is the key question addressed in our investigation.

 While a purely physics based function for the spatial variation in the perturbation may not be possible, this function may possibly be estimated from data. Machine Learning (ML) based approaches are being increasingly applied to fluid dynamics and turbulence applications\cite{duraisamy2019turbulence, ihme2022combustion, brunton2020machine}. Some ML approaches have used data to develop functions that can predict the discrepancy in turbulence model predictions \cite{xiao2016quantifying,wu2018physics,heyse2021estimating,heyse2021data,zeng2022adaptive}. However many of these Machine Learning models are complex and are prone to overfitting. This overfitting limits the generalizability of the trained models and they end up being accurate only for flows that they are trained on. The complex nature of the machine learning models also necessitates large amounts of relevant data for training them. In engineering design this is not always possible especially when new designs are being considered. The complex machine learning models are also prone to being black box models where their inner working is not well understood. This limits our ability for Verification and Validation of such models leading to issues of trust and with adoption of these ML models in engineering applications. Most of the prior investigations utilizing machine learning approaches for uncertainty quantification of turbulence models have used polynomial regression\cite{chu2022model}, Probabilistic Graphical Models (PGMs), Random Forests, or Fully Connected Neural Networks (FCNNs). These model approaches ignore the non-locality of turbulence physics where the evolution of a turbulent flow is affected by far off in the flow domain too. Convolutional Neural Networks (CNNs) can bring in non-locality in the models used to assess turbulence model discrepancy. Augmenting the ML approach with this domain knowledge can lead to an inductive bias in the learning framework that leads to improved performance with lesser data. Moreover we also select our CNN model hyperparameters so as to limit the need for data. In this investigation we utilize a computationally inexpensive machine learning approach that attempts to use this non-local information via CNNs. 

In summary, in this investigation we utilize a novel approach that attempts to account for the missing non-local physics in eddy viscosity models by the use of Convolutional Neural Networks. We use this approach to develop a function that can predict the magnitude of perturbations required in the application of the Eigenspace Perturbation Method. This is an important need for turbulence modeling and engineering design. This can lead to more efficient designs that maintain safety and reliability but reduce the factors of safety used in traditional engineering design protocols.

\section{Methodology}
In this section we shall provide the background for the methodology used in this study. We start with an overview of the Eigenspace Perturbation Method, outlining its strengths and limitations. Then we discuss our prescribed correction involving a CNN based marker function that modulates the degree of the eigenvalue perturbation. 

\subsection{Eigenspace Perturbation}
Eddy Viscosity based models use the concept of a turbulent or eddy viscosity for closure of the evolution equation for the Reynolds stresses. This is also referred to as the Boussinesq Hypothesis \cite{pope2001turbulent}. The instantaneous value of the Reynolds Stresses is assumed to be linearly proportional to the instantaneous value of the mean rate of strain tensor as

\begin{equation}
    \left\langle u_i u_j\right\rangle=\frac{2}{3} k \delta_{i j}-2 v_{\mathrm{t}}\left\langle S_{i j}\right\rangle.
\end{equation}

Here $k$ denotes the turbulence kinetic energy, $\delta_{i j}$ is the Kronecker delta tensor, $\nu_{t}$ is the eddy viscosity coefficient, and $\left\langle S_{i j}\right\rangle$ is the mean rate of strain. This assumption greatly reduces the complexity of the RANS equations and makes engineering simulations simpler. But this assumption is also very limiting and makes eddy viscosity models inaccurate for complex flows. For example the linear relationship implies that the principal co-ordinates of the mean rate of strain tensor and the Reynolds stress tensor are identical. This is not correct for flows with turbulent separation or re-attachment. This also ignores any effects of the mean rate of rotation on the evolution of the Reynolds stresses, limiting its utility in cases with significant streamline curvature or rotation dominated flows.  

To estimate the uncertainties due to such modeling simplifications the Eigenspace Perturbation Method \cite{iaccarino2017eigenspace} introduces perturbations in the spectral representation of the model predictions of the Reynolds stress tensor

\begin{equation}\label{Eq:Rij_perturb}
        \left\langle u_{i} u_{j}\right\rangle^{*}=2 k^{*}\left(\frac{1}{3} \delta_{i j}+v_{i n}^{*} \hat{b}_{n l}^{*} v_{j l}^{*}\right),
\end{equation}

where $\hat{b}_{k l}^{*}$ is the perturbed eigenvalue matrix, $v_{i j}^{*}$ is perturbed eigenvector matrix, $k^{*}$ is the perturbed turbulence kinetic energy. From a modeling perspective the EPM replaces the linear, isotropic eddy viscosity assumption with the general relationship in which the eddy viscosity is a fourth order tensor that incorporates the anisotropic nature of turbulent flows\cite{mishra2019theoretical}.

Due to its physics based nature, the EPM weighs all physically possible outcomes equally leading to overly conservative uncertainty estimates in many cases. In many cases this leads to overly conservative and uncalibrated uncertainty estimates. In a nutshell while the EPM provides a solution for ``how to perturb" it does not address the question of ``how much to perturb". This degree of perturbation is often left as a user defined parameter. For new and different designs, the user may have very limited knowledge to infer such parameters. This degree The key limitation of the EPM is to estimate the degree and extent of the introduced perturbation and its variation over the flow domain. This is the key question addressed in the present investigation.


\subsection{Corrective Function for RANS Predictions}
The highest fidelity of turbulence simulations involve Direct Numerical Simulation (DNS), where all the scales of turbulence are resolved in simulation. RANS simulations on the other hand use closure models for all the scales. Therefore DNS simulations can be much more accurate than RANS simulations. On the other hand RANS simulations are computationally less expensive. We outline a correction term that reduces the discrepancy between the lower fidelity RANS predictions and the high fidelity DNS results. This is referred to as the Correction function. Prior studies have also introduced correction functions and terms for improving RANS predictions, such as a linear correction function \cite{ahlfeld2016multi} and additive and multiplicative versions of correction functions\cite{voet2021hybrid}. 

 For both the RANS and DNS simulation, we can summarize their results as the function of the perturbed turbulence kinetic energy $k^{*}$: 
 
\begin{equation}\label{Eq:Marker_Mk_Method}
    k^{*} = f(x,y)
\end{equation}

where $x$ and $y$ are coordinates in a two-dimensional computational domain, and $f$ is the mapping from every coordinate $(x, y)$ to $k^{*}$, embedded in triples $(x, y, k^{*})$ from simulation results.

Without assuming a specific form, the correction function for RANS is a mapping between two functions:

\begin{equation}
Z: f^{\mathrm{RANS}}(x,y) \rightarrow f^{\mathrm{DNS}}(x,y)
\end{equation}

with $k^{\mathrm{DNS}} = f^{\mathrm{DNS}}(x, y)$ and $k^{\mathrm{RANS}} = f^{\mathrm{RANS}}(x, y)$, we can rewrite $Z$ as a mapping $\zeta$ between points that comprises $f^{\mathrm{RANS}}$ and $f^{\mathrm{DNS}}$

\begin{equation}
\zeta: (x, y, k^{\mathrm{RANS}}) \rightarrow (x, y, k^{\mathrm{DNS}})
\end{equation}

Consider the model error for RANS and DNS in terms of kinetic energy, we have

\begin{equation}
   p^{\text {RANS }}\left(K_g \mid x, y\right)=p\left(k_g=k^{\text {RANS }} \mid x, y\right)
\end{equation}

\begin{equation}
    p^{\text {DNS }}\left(K_g \mid x, y\right)=p\left(k_g=k^{\text {DNS }} \mid x, y\right)
\end{equation}

where $K_g$ is the unknown ground truth of kinetic energy at $(x, y)$.

Kinetic energy resulted from DNS simulation results $p^{\mathrm{RANS}}$ can be estimated with kinetic energy from RANS simulation $p^{\mathrm{DNS}}$ and its correction function $g$ as

\begin{equation}
p^{\mathrm{DNS}}\left(K_g \mid x, y\right)=g\left(k^{\mathrm{RANS}}, x, y\right) p\left(k^{\mathrm{RANS}} \mid x, y\right)
\end{equation}

Because $k^{\mathrm{DNS}} = f^{\mathrm{DNS}}(x, y)$ and $k^{\mathrm{RANS}} = f^{\mathrm{RANS}}(x, y)$,
at each $x$, we have that $k_x^{\mathrm{DNS}} = f_x^{\mathrm{DNS}}(y)$ and $k_x^{\mathrm{RANS}} = f_x^{\mathrm{RANS}}(y)$, assuming both $f_x^{\mathrm{RANS}}$ and $f_x^{\mathrm{DNS}}$ are continuous, that is, $\forall \epsilon > 0,\allowbreak\, \exists \delta > 0,\allowbreak\, s.t.\, \forall \abs{d} < \delta, \allowbreak \, \abs{f_x(y + d) - f_x(y)} < \epsilon$. We can approximate $g(k^{\mathrm{RANS}}, x, y)$ with $\hat{g}(\mathbf{k}_{x,y,\delta}^{\mathrm{RANS}})$, where $\mathbf{k}_{x,y,\delta}^{\mathrm{RANS}} = [k_{x, y_0}^{\mathrm{RANS}}, k_{x, y_1}^{\mathrm{RANS}}, \cdots]^\top$ and $y_0, y_1, \dots \in [y - \delta, y + \delta]$. In other words, we can learn $\hat{g}$ with paired $(\mathbf{k}_{x,y,\delta}^{\mathrm{RANS}}, \mathbf{k}_{x,y,\delta}^{\mathrm{DNS}})$.

In this study, the DNS are performed using normalized variables namely $U_{0} =1$ and $H = 1$. The low-fidelity RANS simulations and high-fidelity DNS are performed at a Reynolds number (Re) of 5600, based on the bulk velocity through the channel $U_{b}  =2.9714 m/s$; a characteristic length hill height = 0.028 m. Here $H$ is the semi-channel height and $U_{0}$ is the center line velocity.

\subsection{Convolutional Neural Network-based Correction Function}
\label{sec:methodology-cnn}


We employed a one-dimensional convolutional neural network (1D-CNN) to learn the correction function $\hat{g}$ from paired RANS and DNS simulation estimated kinetic energy $(\mathbf{k}_{x,y,\delta}^{\mathrm{RANS}}, \mathbf{k}_{x,y,\delta}^{\mathrm{DNS}})$. Because our approximated correction function $\hat{g}$ only depends on the neighbor of $k^{\mathrm{RANS}}$ and coordinates $(x, y)$ are only used to group neighbors of $k^{\mathrm{RANS}}$, we grouped simulation data by $x$ and transformed $(y, k)$ at $x$ into $\mathbf{k}_{x,y,\delta}^{\mathrm{RANS}}$ via a rolling window parameterized by window size. Our 1D-CNN has four-layers and in total 86 parameters: a single model for all zones at any $x$ to correct RANS towards DNS.

\begin{figure}[h]
    \centering
    \includegraphics[width=\linewidth]{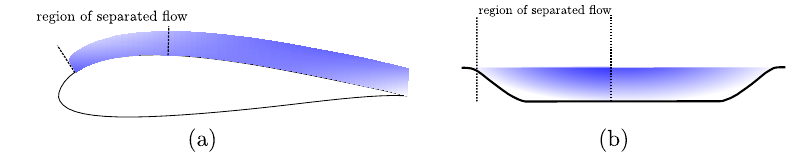}
    \caption{(a) flow passing over an SD7003 airfoil and (b) flow over a two-dimensional periodically arranged hills.} 
    \label{fig:Schematics_SD7003_hills.pdf}
\end{figure}

\section{Experiment Setup and Data Sources}

We experimented our lightweight CNN-based approach to approximate the correction function for RANS on two datasets: an in-house RANS/DNS \cite{zhang2021turbulent,chu2022model} dataset and the public RANS/DNS dataset \cite{voet2021hybrid}. The in-house RANS/DNS dataset \cite{chu2022model,zhang2021turbulent} was obtained by considering the flow around an SD7003 airfoil, as shown in Fig. \ref{fig:Schematics_SD7003_hills.pdf} (a). On the suction side, a separation bubble is formed due to the adverse pressure gradient. The public RANS/DNS dataset \cite{voet2021hybrid} was generated from the two-dimensional channel flow over periodically arranged hills, as shown in Fig. \ref{fig:Schematics_SD7003_hills.pdf} (b). Similar to the flow over the airfoil, the flow also experiences adverse pressure gradient when encountering the curved surface of the hill. This causes the formation of a separation bubble behind the hill.

\subsection{The In-house RANS/DNS Dataset}
\begin{figure*}[h!]
    \centering
    \includegraphics[width=\linewidth]{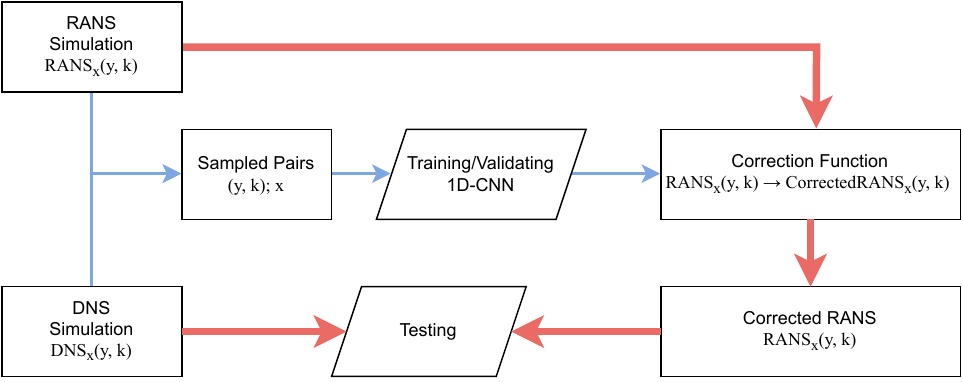}
    \caption{Data-flow Diagram for Experiments. Blue path is the training/validating path, and the red path is the validation path. Blue path represents small amount (about 20\%) of data through flow. Red path represents large amount (about 80\%) of data through flow.} 
    \label{fig:data-flow.pdf}
\end{figure*}

The present study uses two in-house RANS and DNS datasets of \cite{chu2022quantification, chu2022model} and \cite{zhang2021turbulent}. These datasets were obtained for the flow over an SD7003 airfoil, where a separation bubble evolves on the suction side. From \cref{fig:data-flow.pdf}, we split $x$-coordinate grouped pairs of $(\mathbf{k}_{x,y,\delta}^{\mathrm{RANS}}, \mathbf{k}_{x,y,\delta}^{\mathrm{DNS}})$ into a training set and a validation set by their group key $x$. For both the in-house datasets, we choose $x$ at only four positions on the geometry of all paired $x$ values. These positions span across the separated region. For the in-house dataset based on the SD7003 airfoil geometry, $x/c = 0.17, 0.25, 0.32, 0.44$ where $c$ is the cord length; for the public dataset based on the two-dimensional periodically arranged hills, $x/H = 0, 0.035, 1.961, 4.885, 6.847$, where $H$ is characteristic length hill height. For each dataset, we use a 80\%--20\% split as training--testing dataset. 

\subsection{The Voet Data Set}
The public dataset of Voet \textit{et al.} \cite{voet2021hybrid} was used to train our CNN model. Those public DNS data are provided in two formats, i.e., \texttt{.dat} and \texttt{.vtr}. We chose the \texttt{.dat} format due to its ease of use. Specifically, the data 
stored in the \texttt{rms\_files1.dat} files are used. However, no names are assigned to columns in the \texttt{rms\_files1.dat} files. The correct column names can be determined using the \texttt{rms\_1.vtr} file, as it contains the data identical to the corresponding \texttt{rms\_files1.dat} file and assigns names to the columns. The correct column names can be obtained by comparing with the same range to the \texttt{rms\_1.vtr} file. The restored column names from left to right in \texttt{rms\_files1.dat} file are \texttt{x}, \texttt{y}, \texttt{uumean}, \texttt{vvmean}, \texttt{wwmean}, and \texttt{p}, respectively. Physically they represent $x$ coordinate, $y$ coordinate, time-averaged turbulent velocity component in $x$, $y$, and $z$ directions. As the result, the DNS data of \texttt{x\_dns, y\_dns, k\_dns} are used in training our CNN model, i.e., \texttt{x\_dns = x}, \texttt{y\_dns = y}, and \texttt{k\_dns = 1/2 (uumean, vvmean, wwmean)}.

On the other hand, the public RANS data are stored in files with the name pattern of \texttt{Results\_bump(\textbackslash d+).csv}, where \texttt{(\textbackslash d+)} is the case index written in the syntax of regular-expression. The column names are clearly given: \texttt{"Velocity[i] (m/s)"} as \texttt{x}, \texttt{"Velocity[j] (m/s)"} as \texttt{y}, and \texttt{"turbulence kinetic Energy (J/kg)"} as \texttt{k}. 

According to Voet \textit{et al.} \cite{voet2021hybrid}, both the high-fidelity DNS and the low-fidelity RANS simulations should have their $x$ and $y$ normalized by the characteristic length hill height $H = 0.028m$. Therefore, the data format of \texttt{x\_{$RANS \textbackslash DNS$}} and \texttt{y\_{$RANS \textbackslash DNS$}} is used in training our CNN model. Besides, Voet \textit{et al.} \cite{voet2021hybrid} defined two main parameters $\alpha$ and $\gamma$ to influence the flow over the periodically arranged hills. Of these two parameters, $\alpha$ changes the hill steepness, and $\gamma$ changes the successive spacing of the hills. Further, the reference test case is defined by $\alpha = 1$ and $\gamma = 1$. Varying these two parameters results in multiple flow cases. It should be noted that the DNS and RANS cases are not paired by case index; instead corresponding DNS and RANS cases can only be determined using $\alpha$ and $\gamma$. We compared these two parameters across DNS cases and RANS cases, and found that DNS Case1 is matched with RANS Case18, similarly, DNS Case2 with RANS Case6, DNS Case3 with RANS Case16, DNS Case4 with RANS Case13, DNS Case5 with RANS Case12, DNS Case6 with RANS Case24, DNS Case7 with RANS Case30.

For both datasets, we validated our trained 1D-CNN by comparing the L1 loss of RANS, denoted $L^1_c(\texttt{rans}) = \abs{ CF^{\mathrm{RANS}}_{k} - CF^{\mathrm{DNS}}_{k} }$, with the L1 loss of 1D-CNN corrected RANS, denoted $L^1_c(\texttt{pred}) = \abs{ CF^{\mathrm{CNN}}_{k} - CF^{\mathrm{DNS}}_{k} }$.

Furthermore, we examined the generalizability of our lightweight CNN-based correction function trained on the public RANS/DNS dataset \cite{voet2021hybrid} by using the correction function to correct RANS for other cases. 

\section{Results and discussion}
\begin{figure} 
    \centering
    \includegraphics[width=\linewidth]{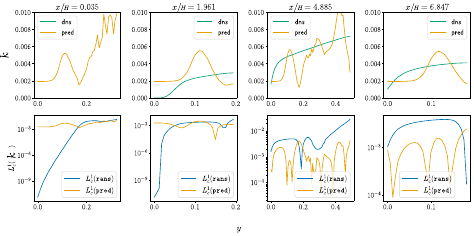}
    \caption{Results for periodically arranged hills. DNS case2 based CNN prediction for turbulence kinetic energy. First row: corrected DNS case 2 (pred) compared with ground truth (dns). For DNS case 2: $\alpha = 0.8$ and $\gamma = 1.0$; DNS case 7: $\alpha = 1.2$ and $\gamma = 1.0$. Second row: Validation of 1D-CNN by comparing L1 loss between $L^1_c(\texttt{rans})$ and $L^1_c(\texttt{pred})$. }
    \label{fig:case2-based-dns2}
\end{figure}

\begin{figure}[H]
    \centering
    \includegraphics[width=\linewidth]{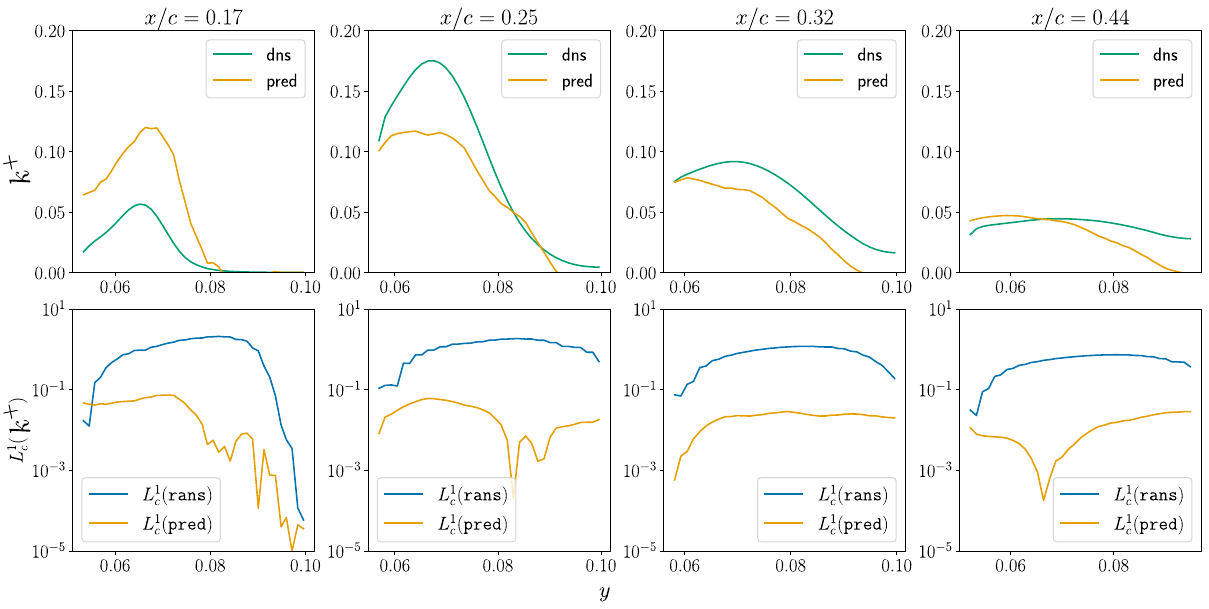}
    \caption{Results for Selig-Donovan 7003 airfoil. DNS-based CNN prediction for normalized turbulence kinetic energy. First row: CNN corrected DNS (\texttt{pred}) compared with ground truth (\texttt{dns}). Second row: Validation of 1D-CNN by comparing L1 loss between $L^1_c(\texttt{rans})$ and $L^1_c(\texttt{pred})$.}
    \label{fig:rans-predicted-dns-main4-viz-loss.pdf}
\end{figure}

Our CNN-based correction function is validated at all paired $x$ locations. \Cref{fig:case2-based-dns2,fig:rans-predicted-dns-main4-viz-loss.pdf} show the results for flow over 2D  periodically arranged hills and flow over an the region of separated flow. From \Cref{fig:case2-based-dns2,fig:rans-predicted-dns-main4-viz-loss.pdf}, the series of CNN predicted DNS profiles in the first row are then smoothed with the moving average (the average of points in a sliding window) with a window size of six consecutive estimations. Our CNN-based prediction for turbulence kinetic energy profile resembles the ground truth DNS despite being trained with only a few pairs of RANS and DNS results.  For sake of simplicity the turbulence kinetic energy is dimensional for \Cref{fig:case2-based-dns2}, while normalized turbulence kinetic is used for \Cref{fig:rans-predicted-dns-main4-viz-loss.pdf}, i.e., $k^{+} = k/U_{\infty}^2$ where $U_{\infty}$ is freestream velocity.


In the present paper, we also used the public RANS/DNS dataset \cite{voet2021hybrid} to train our CNN-based model. In Fig. \ref{fig:case2-based-dns2}, our CNN-based model being trained based on the case two of the public RANS/DNS dataset \cite{voet2021hybrid} is used to predict the $k$ profiles for DNS case seven, with the corresponding $\alpha = 0.8$ and $\gamma = 1.0$ for case 2 and $\alpha = 1.2$ and $\gamma = 1.0$ for case 7. turbulence kinetic energy is not normalized as it does not affect our performance of our CNN model. Among these four locations, the CNN-based prediction for turbulence kinetic energy profile shows a relatively large discrepancy at $x/H = 0.035$ where a high level of chaos is present due to the intense separation, although the $L^1_c(\texttt{pred})$ is significantly reduced by around six orders of magnitude in the near-wall region compared to $L^1_c(\texttt{rans})$. While as the flows moves further downstream and the degree of separation reduces, our CNN-based predictions show a tendency of approaching to the ground truth DNS, although the CNN-based predictions fluctuate around the ground truth DNS. The $L^1_c(\texttt{pred})$ error in the second row is basically at the same order of magnitude for $x/c = 0.035$ and $x/c = 1.961$ except in the region next to the wall $L^1_c(\texttt{pred})$ lying significantly above $L^1_c(\texttt{rans})$. On the other hand, the magnitude of $L^1_c(\texttt{pred})$ error for $x/H = 4.885$ and $x/H = 6.847$ overall sits below that of $L^1_c(\texttt{rans})$ at approximately one order of magnitude. 

For both cases, our CNN-based predictions for $k$ profiles overall sit close to the DNS data at any $x$ location, i.e., the discrepancy in general reduces as the flow proceeds further downstream. This is clearly reflected from the computed $L^1$ error in the second row of \Cref{fig:rans-predicted-dns-main4-viz-loss.pdf}, i.e., a $L^1_c(\texttt{pred})$ drop of two orders compared to $L^1_c(\texttt{rans})$. From \Cref{fig:rans-predicted-dns-main4-viz-loss.pdf}, it is interesting to note that the CNN-based predictions for $k$ profiles tend to approach closer to the DNS profile as the flow proceeds further downstream. At $x/c = 0.17$ the flow experiences rather intense separation and flow features are complex, the discrepancy between the CNN-based prediction and ground truth DNS for $k$ profile is relatively large. It must be emphasized that as the flow moves further downstream the degree of separation reduces and the CNN-based model tends to become more trustworthy within the region of separated flow. It indicates that the generalizability of our CNN-based model reduces when dealing with complex flows, e.g., where rather complex flow features evolve.

A key source of limitation of data driven ML models is their inability to generalize from their training dataset to flows that may be different from the training data. For our experiments, it can be concluded that our CNN-model is relatively robust and retains a consistent behavior even with different training data sets. Overall our model can give satisfactory results within the separated region, in particular the aft portion of the separated region.


\section{Discussion}

We proposed a CNN approach to approximate the correction function that corrects RANS simulation towards DNS simulation. We further examined our method on two datasets: 
\begin{enumerate}
    \item A flow over an SD7003 airfoil at $8^\circ$ angle of attack and the Reynolds number based on the cord length of $Re_{c} = 60000$ \cite{zhang2021turbulent}. A laminar separation bubble evolves on the suction side of the airfoil whereby the flow undergoes transition to turbulence,

    \item the second flow case is generated from the DNS two-dimensional channel flow over periodically arranged hills \cite{voet2021hybrid}. The flow experiences adverse pressure gradient when encountering the curved surface of the hill.
\end{enumerate}

It should be noted that a separation bubble occurs for both datasets. The RANS results deviate from the DNS data in both flow scenarios and our CNN-based correction function can significantly reduce the $L^1$ error of RANS- from DNS-simulations.


For both datasets, the CNN-based correction function is trained on paired RANS-DNS simulated turbulence kinetic energy using less than $20\%$ positions along $x$-axis, while the CNN-based function is still accurate for the remaining $80\%$ positions. In other words, our CNN-based correction can be used to predict RANS-DNS simulated turbulence kinetic energy at any $x$ coordinate with only a fraction of the whole $x$ coordinates. Furthermore, our lightweight CNN model uses the $y$ coordinates for grouping RANS-simulated turbulence kinetic energy within a neighborhood. The results of our CNN-based correction function suggests that RANS results might be improved by leveraging information embedded in the positions within a close neighbor, which is independent of the absolute coordinates $(x, y)$. The lightweight CNN-based correction function trained on one case can still help smooth and reduce the error of RANS-based results for other cases.

There are relatively few studies for correcting the perturbed turbulence kinetic energy. Very recently, the study of Chu \textit{et al.} \cite{chu2022model} assessed the effect of polynomial regression on the estimation of the perturbed turbulence kinetic energy. Our CNN-based correction method has readily implications on practical applications, such as, to be coupled to the eigenspace perturbation approach of Emory \textit{et al.} \cite{emory2013modeling}. The eigenspace perturbation approach has been implemented within the OpenFOAM framework to construct a marker function for the perturbed turbulence kinetic energy \cite{chu2022model}. Our CNN-based correction method can be used as a new marker function to predict the perturbed turbulence kinetic energy. 

\subsection{Application of the lightweight CNN-based correction function on UQ for an SD 7003 airfoil}
Our CNN-based correction function method can be applied to different flow cases to correct RANS towards DNS. In this section, the CNN-based correction function is applied to the SD7003 airfoil case to predict the perturbed turbulence kinetic energy.

\begin{figure*}[t]
         \centering
         \includegraphics[width=\textwidth, trim={2.4mm 0 4mm 0},clip]{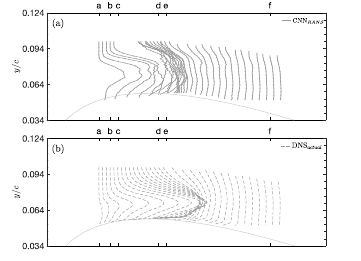}
        \caption{(a) CNN corrected RANS (\texttt{CNN\_{DNS}}) {(solid-dotted lines)} of the normalized perturbed turbulence kinetic energy and (b) ground truth (\texttt{CNN\_{RANS}}) along the suction side of the SD7003 airfoil (geometry depicted by gray line): from left to right are zone $ab$, zone $cd$ and zone $ef$. There are 32 positions on the suction side of the airfoil.}
        \label{fig:CNN_DNS.pdf}
\end{figure*}

The CNN corrected RANS and ground truth profiles for the turbulence kinetic energy normalized with the freestream velocity squared, $k^{*}/U_{\infty}^2$ and $k/U_{\infty}^2$ are shown in Figs. \ref{fig:CNN_DNS.pdf} (a) and (b), respectively. The $k^{*}/U_{\infty}^2$ and $k/U_{\infty}^2$ profiles are equally spaced for the $ab$ and $cd$ zone with $x/c = 0.01$, and a uniform spacing of $x/c = 0.02$ is used for the $ef$ zone. It is clear that the $k^{*}/U_{\infty}^2$ and $k/U_{\infty}^2$ profiles are more densely packed for the $ab$ and $cd$ zone, within which the flow features are complex due to the presence of separation and reattachment. 

From Figs. \ref{fig:CNN_DNS.pdf} (a) and (b), the CNN corrected DNS profiles in general exhibit a similar trend as that for the ground truth dataset, as both profiles show a gradual increase in the $ab$ and $cd$ zone. Then a reduction of the profile is observed further downstream in the $ef$ zone. 

In Fig. \ref{fig:CNN_DNS.pdf} (a), CNN corrected RANS profiles in general increase in magnitude as the flow moves further downstream, which is qualitatively similar to the ground truth profiles. Further, it should be noted that the CNN corrected RANS profiles increase in a somewhat larger magnitude than that for the ground truth in the $ab$ zone. The discrepancy is more than $50\%$ at the beginning of the $ab$ zone and gradually reduces as the flow moves further downstream, which indicates that a better accuracy of our CNN model is yielded further downstream. This behavior becomes more clear for the $cd$ and $ef$ zone. In the region where the end of the $cd$ zone meets the beginning of the $ef$ zone, the ground truth profiles are clustered due to the complex flow feature of the reattachment \cite{chu2022quantification}, as shown in Figs \ref{fig:CNN_DNS.pdf} (b). This clustering behavior is successfully captured by our CNN model, as shown in Fig. \ref{fig:CNN_DNS.pdf} (a). In the $ef$ zone, our CNN model gives overall accurate predictions for the $k^{*}/U_{\infty}^{2}$ profiles, i.e., the CNN corrected RANS profiles and the ground truth profiles are almost identical.



\section{Conclusion}
\label{sec:Conclusion}

While researchers have attempted to use ML models to augment the EPM\cite{heyse2021data, heyse2021estimating, matha2023evaluation}, this investigation is the first to examine the projection from RANS to DNS using the CNN approach. The use of a convolution filter enables the mapping to include non-local information whose absence is one of the key sources of model limitation in single-point RANS closures. Our experiment results suggest that the CNN-based correction function corrects the RANS predictions for perturbed turbulence kinetic energy towards the in-house DNS data. A projection that can approximate the in-house DNS data reasonably well from RANS might exist independent of $x$. Our methodology can be easily extended to analyze flows over different types of airfoils.

Our findings are subject to following limitations: our CNN model was trained with only two datasets. Future work may include validation with more datasets using different flow cases, e.g., different types of airfoils.  In addition, our CNN-based correction function will be integrated with the eigenspace perturbation framework to result in accurate perturbations and hence improved estimation of RANS UQ. Finally, we are investigating the trained CNN using ML interpretability approaches to verify if the model has learned physics based feature maps. This understanding would enable a higher degree of trust in this physics based machine learning approach.

\begin{acknowledgments}
The author Minghan Chu thanks the author Weicheng Qian for his tremendous help in training the machine learning model.
\end{acknowledgments}


\section*{Data Availability Statement}
The data that support the findings of this study are available from the corresponding author upon reasonable request. 

\appendix

\bibliography{aipsamp}

\end{document}